\documentclass[preprint,final,5p,times,twocolumn]{elsarticle}

\usepackage[colorlinks,linkcolor=blue,anchorcolor=blue,citecolor=blue,urlcolor=blue]{hyperref}
\usepackage{graphicx}
\usepackage{epstopdf}
\usepackage{dcolumn}
\usepackage{textcomp}
\usepackage{gensymb}
\usepackage{booktabs}
\usepackage{amsmath}
\usepackage{amssymb}
\usepackage{color}
\usepackage{hyperref}
\usepackage{url}
\usepackage{siunitx}
\usepackage{subfigure}
\usepackage{longtable}
\usepackage{array}
\usepackage{booktabs}
\usepackage{verbatim}
\usepackage{threeparttable}
\usepackage{orcidlink}
\usepackage{dcolumn}
\usepackage[switch]{lineno} 

\begin{document}
\begin{frontmatter}

\title{Development and Characterization of a High-Resolution and High-Sensitivity Collinear Resonance Ionization Spectroscopy Setup}

\author[PKU]{H.~R.~Hu}
\author[PKU]{Y.~F.~Guo}
\author[PKU]{X.~F.~Yang\orcidlink{0000-0002-1633-4000}\corref{cor1}}\ead{xiaofei.yang@pku.edu.cn}
\author[PKU]{Z.~Yan}
\author[PKU]{W.~C.~Mei}
\author[PKU]{S.~J.~Chen}
\author[PKU]{Y.~S.~Liu}
\author[PKU]{P.~Zhang}
\author[PKU]{S.~W.~Bai\orcidlink{0009-0001-2322-570X}}
\author[PKU]{D.~Y.~Chen}
\author[PKU]{Y.~C.~Liu}
\author[PKU]{S.~J.~Wang}
\author[PKU]{Q.~T.~Li}
\author[PKU]{Y.~L.~Ye\orcidlink{0000-0001-8938-9152}}
\author[CIAE]{C.~Y.~He}
\author[IMP]{J.~Yang}
\author[LZU]{Z.~Y.~Liu}

\cortext[cor1]{Corresponding author}
\address[PKU]{State Key Laboratory of Nuclear Physics and Technology, School of Physics, Peking University, Beijing 100871, China}
\address[CIAE]{China Institute of Atomic Energy (CIAE), P.O. Box 275(10), Beijing 102413, China}
\address[IMP]{Institute of Modern Physics, Chinese Academy of Sciences, Lanzhou 730000, China}
\address[LZU]{School of Nuclear Science and Technology and Frontiers Science Center for Rare Isotopes, Lanzhou University, Lanzhou 730000, China}

\begin{abstract}
With the recent implementation of a radio-frequency quadrupole (RFQ) cooler-buncher and a multi-step laser resonance ionization technique, our previously developed collinear laser spectroscopy setup has been successfully upgraded into a fully functional collinear resonance ionization spectroscopy system. The new system was fully characterized using a bunched ion beam at 30~keV, during which
hyperfine structure spectra of $^{85,87}$Rb isotopes were measured. An overall efficiency exceeding 1:200 (one resonant ion detected for every 200 ions after the RFQ cooler-buncher) was achieved while maintaining a spectral resolution of 100 MHz. Under these conditions, the extracted hyperfine structure parameters and isotope shift for $^{85,87}$Rb show excellent agreement with the literature values. These results demonstrate the system's capability to perform high-resolution and high-sensitivity laser spectroscopy of neutron-rich Rb isotopes, which are expected to be produced at the Beijing Radioactive Ion-beam Facility at a rate of approximately 100 particles per second.

\end{abstract}

\end{frontmatter}

\begin{figure*}[!t]
\center
\includegraphics[width=0.95\hsize]{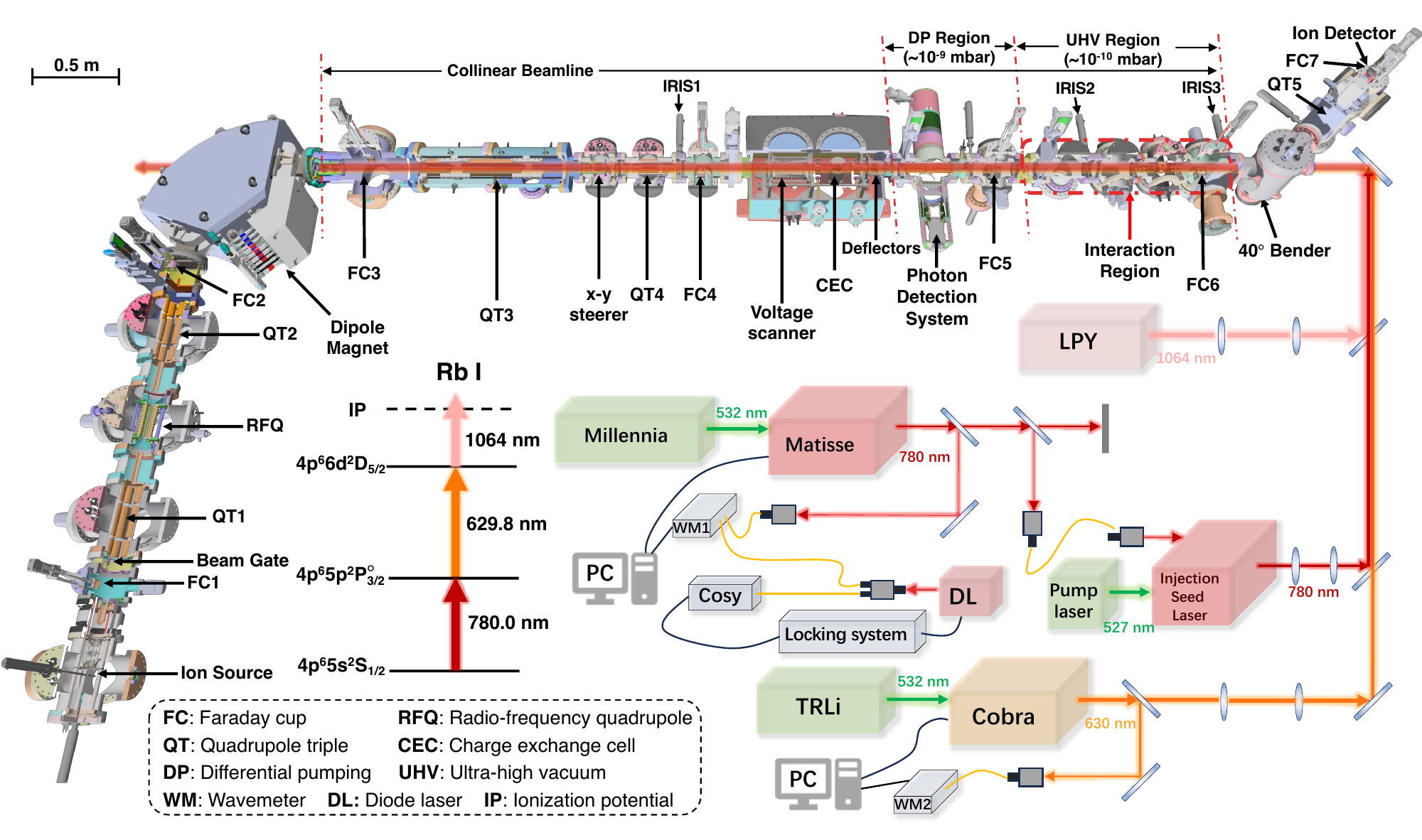} 
\vspace{-2mm}
\caption{A schematic view of the entire PLASEN system, including the offline ion source, the RFQ cooler-buncher, the dipole magnet, collinear resonance ionization laser spectroscopy, the laser optical system, and the ionization scheme of Rb I used for this test experiment. }
\vspace{-2mm}
\label{Fig1}
\end{figure*}

Understanding the exotic structure of unstable nuclei is a key focus in nuclear physics, as it is crucial for gaining deeper insights into nuclear forces and behavior of microscopic quantum many-body systems~\cite{NRP2025}.The basic properties of unstable nuclei are strongly related to nuclear structure and nucleon-nucleon interaction, making them essential for exploring various novel phenomena that emerge in exotic nuclei. Laser spectroscopy is one of the most powerful techniques for investigating nuclear properties. It enables nuclear-model-independent measurements of nuclear spins, magnetic and quadrupole moments, and charge radii, by probing the hyperfine structure~(HFS) and isotope shift~(IS) of the corresponding atoms, ions or even molecules~\cite{PPNP2023}. However, HFS effects and IS contribute only about one part in a million of the total transition frequency, necessitating highly precise measurement techniques. 
\par Collinear laser spectroscopy~(CLS) using laser-induced fluorescence (LIF) detection, since its development in the 1970s, has made irreplaceable contributions to the study of nuclear properties and exotic nuclear structure, owing to its advantage in achieving high-resolution spectroscopy measurement~\cite{COLLAPS2017}. In particular, the combination of a radio-frequency quadrupole (RFQ) cooler-buncher with the CLS technique has significantly enhanced measurement sensitivity, typically down to a few thousand particles per second~(pps)~\cite{COLLAPS2017}. With certain variant--such as decay detection following optical pumping and state-dependent neutralization~\cite{roc}--the CLS technique can even achieve record-level sensitivity, although it is applicable only to a limited number of specific elements.
\par With the increasing production and experimental accessibility of unstable nuclei at radioactive beam~(RIB) facilities worldwide, the past two decades have witnessed a renaissance in the development and upgrading of laser spectroscopy techniques~\cite{PPNP2023}, aiming at achieving higher sensitivity. This advancement is driven by the growing interest in exploring exotic nuclei farther from $\beta$-stability valley, which however present significant experimental challenges due to their shorter lifetimes, lower production rates, and pronounced isobaric contamination. This progress has led to the emergence of several novel techniques, such as the collinear resonance ionization spectroscopy (CRIS) setup at ISOLDE-CERN~\cite{CRIS-NIMB2020}, which is potentially applicable to a wide range of isotopes across the nuclear chart; the Multi Ion Reflection Apparatus for Collinear Laser Spectroscopy (MIRACLS) at ISOLDE-CERN~\cite{MIRACLS}, optimized for selected elements; laser ionization spectroscopy in a supersonic gas jet for isotopes of heavy mass region~\cite{Ac-moment2017}; and RAdiation-Detected Resonance Ionization Spectroscopy (RADRIS) for the studies of superheavy element~\cite{RADRIS2023}. 
\par Among these, CRIS technique has attracted growing attention due to its ability to simultaneously achieve high resolution and high sensitivity. It has proven effective for investigating nuclear properties across a broad range of isotopes~\cite{Cu2017,PPNP2023}, for decay spectroscopy of purified isomeric states~\cite{Fr-PRX}, and for pioneering studies of radioactive molecules~\cite{RaF}. By performing resonance ionization spectroscopy(RIS) with a fast ion beam overlapped collinearly or anticollinearly with multiple pulsed laser beams, the CRIS method significantly enhances measurement sensitivity while preserving the high spectral resolution characteristic of standard CLS~\cite{COLLAPS2017}. This makes it particularly well-suited for studying exotic nuclei across diverse regions of the nuclear chart.
\par To investigate nuclear properties of unstable nuclei at existing and forthcoming RIB facilities in China~\cite{BRIF2020,HIAF2022}, we developed a CLS setup employing LIF detection at the first stage~\cite{CLS-PKU,CLS-BRIF}. The CLS setup was successfully commissioned using stable calcium~\cite{CLS-PKU} and unstable potassium at BRIF~\cite{CLS-BRIF}. While characterized by high resolution~\cite{CLS-PKU}, the system has limited sensitivity (typically few thousands pps), making it unsuitable for studying very low intensity RIBs. Furthermore, the energy spread of BRIF-produced RIB poses additional challenge that constrains the achievable spectral resolution~\cite{CLS-BRIF}. Building on this initial development, we upgraded the setup by integrating the RIS technique~\cite{early-CRIS-PKU}. More recently, to address challenges related to spectral resolution and experimental sensitivity arising from energy spread and the continuous operation mode of BRIF's RIBs, respectively~\cite{CLS-BRIF}, we incorporated a RFQ cooler-buncher~\cite{RFQ-PKU}. As a result of these upgrades, we have now established a fully-functional, high-resolution and high-sensitivity collinear resonance ionization laser spectroscopy setup, named PLASEN-Precision LAser Spectroscopy for Exotic Nuclei.
\begin{figure}[!t] 
\center
\includegraphics[width=0.9\hsize]{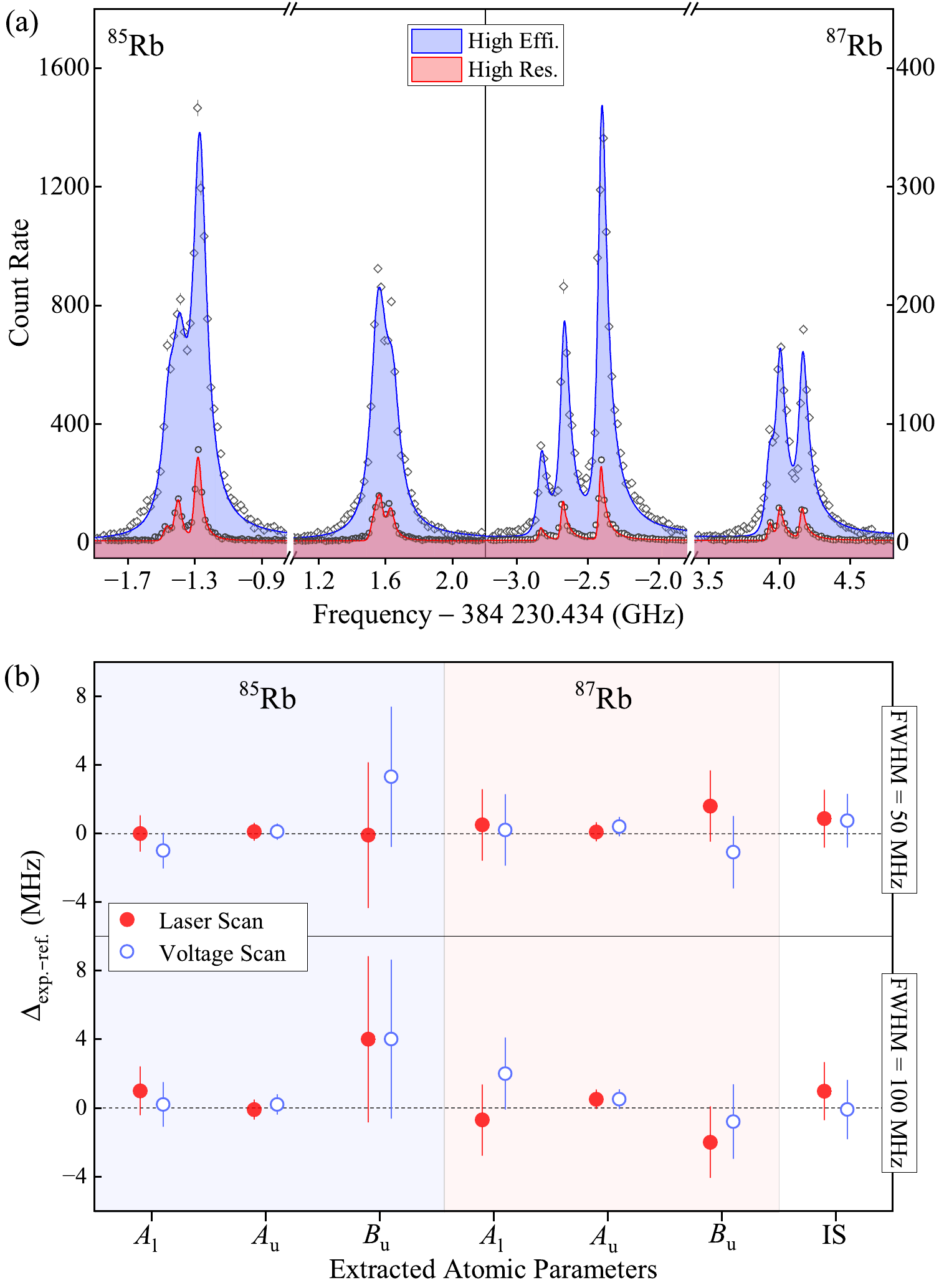} \\
\vspace{-4mm}
\caption{(a)~HFS spectra of $^{85,87}$Rb measured by scanning the laser frequency. Spectra fitted with blue lines correspond to measurement with an efficiency of 1:200 and a resolution 100~MHz, while those fitted with red lines represent measurement with a higher resolution of 50~MHz but a lower efficiency of 1:500. (b)~Differences between magnetic and quadrupole HFS parameters~($A_{\rm u}$, $A_{\rm l}$ and $B_{\rm u}$) of $^{85,87}$Rb isotopes in this work and values from literature~\cite{Rb1981}. Red solid dots and blue open dots represent results from laser frequency scan and voltage scan, respectively. The upper and lower panels show results from HFS spectra with resolution of 50~MHz and 100~MHz, respectively. }
\vspace{-4mm}
\label{Fig2}
\end{figure}
\par Figure~\ref{Fig1} presents a schematic diagram of the PLASEN setup, including the RFQ cooler buncher, the resonance ionization spectroscopy and the laser systems used for Rb HFS measurement. The Rb ions are produced by a surface ion source. After extraction and acceleration to 30~keV, the continuous ion beam is injected into the RFQ via an electrostatic quadrupole triplet lens (QT1). Inside the RFQ, the ion beam is cooled and bunched, forming ion bunches with a temporal width of approximately 2~$\mu$s~\cite{RFQ-PKU} at a period of 10~ms. The RFQ transmission efficiency can reach up to 60\%, as recently demonstrated (see Ref.~\cite{RFQ-PKU} for details). After extraction from the RFQ, the ion bunches are re-accelerated to 30~keV and then refocused by QT2 to improve the beam profile and enhance transmission efficiency. Subsequently, the ion beam is mass-separated and deflected into the collinear beamline by a 90$^{\circ}$ dipole magnet. Placing the magnet downstream of the RFQ is intended for future studies on molecular formation within the RFQ. At 30-keV beam energy, this dipole magnet can achieve a mass resolution ($m/\Delta m$) greater than 200, which is sufficient for separating the heavy isotopes and their Fluoride molecule or Hydroxide molecule. Following the mass separation, the ion beam with a single mass number~(e.g.~$^{85}$Rb) is reshaped, focused, and further transported into the interaction region (IR) through QT3 and QT4 and multiple sets of $x-y$ steerer.
\par In the collinear beamline (Fig.\ref{Fig1}), the Rb ions are firstly neutralized in a charge exchange cell (CEC) filled with high-density sodium vapor. Non-neutralized ions are deflected away via the deflector plates located downstream. The neutral atoms are subsequently delivered into the IR, where atom bunches overlap spatially and synchronize temporally with three pulsed laser beams, facilitating laser resonance ionization. A cone-shaped voltage-scanner installed upstream of the CEC is designed for performing Doppler tuning (voltage scan)~\cite{CLS-BRIF} while minimizing the beam trajectory changes up to the IR during voltage scan. The IR is maintained at ultra-high vacuum (UHV) of approximately $10^{-10}$~mbar to suppress non-resonant collision ionization of the neutral beam by residual gases. This vacuum level is achieved via a differential pumping (DP) region, which quasi-isolates the 10$^{-8}$ mbar vacuum region inside the CEC and the $10^{-10}$~mbar UHV in the IR. A photon detection system is also available downstream of the CEC for measuring the HFS spectra using LIF detection~\cite{CLS-PKU}. This system additionally serves as the DP region, with its two aperture arrays functioning as the DP tubes~\cite{early-CRIS-PKU}. Finally, the laser-resonantly ionized ions within the IR are guided into the MagneTOF ion detector by a 40$^{\circ}$ bender and a quadrupole triplet lens (QT5).
\par As shown in Fig.\ref{Fig1}, a three-step excitation and ionization scheme is employed to measure the HFS spectra of Rb atoms (Rb I). A narrow-band pulsed laser is first used to resonantly excite Rb I through \mbox{$5s$ $^{2\!}S_{1/2}$} $\to$ \mbox{$5p$ $^{2\!}P^{\rm{o}}_{3/2}$} (D2) transition. This laser is produced by an injection-locked titanium-sapphire (Ti:Sa) system~\cite{inseeded2018} at a repetition rate of 100 Hz. This system is seeded by a cw Ti:Sa laser and pumped by a 527-nm pulsed Nd:YLF laser. Long-term laser-frequency stabilization is accomplished using a high-accuracy wavemeter (WM1, High-Finesse WS8) which is calibrated or corrected using a tunable diode laser locked to a HFS transition of $^{87}$Rb in a temperature-controlled vapor cell. The second laser at 629.8 nm, involves the \mbox{$5p$ $^{2\!}P^{\rm{o}}_{3/2}$} $\to$ \mbox{$6d$ $^{2\!}D_{5/2}$} transition, which is produced by a pulsed dye laser (Cobra-Stretch, Sirah Lasertechnik) pumped by a 532-nm Nd:YAG laser at 100 Hz. Its frequency is optimized to maximize the RIS counts and monitored by a second wavemeter (WM2, High-Finesse WS6). The final non-resonant ionization is performed using a 1064-nm Nd:YAG laser. Timing synchronization among all three laser pulses and ion bunches is managed through a series of digital-delay pulse generators.
\par To verify the overall performance and demonstrate the full capabilities of the PLASEN system, we performed collinear resonance ionization spectroscopy experiments on $^{85,87}$Rb isotopes by measuring their HFS spectra through the D2 transition.
\par In this experiment, using the bunched beam from the RFQ, the overall ion beam transmission of the collinear beamline has been significantly improved, consistently greater than 80\% verified by Faraday cups (FC2 to FC7). The Rb ions are neutralized in the CEC with a neutralization efficiency of over 80\%. Three laser beams with 1-cm diameter are introduced anti-collinearly through the collinear beamline (see Fig.~\ref{Fig1}). Their optimal overlap is confirmed using two phosphor screens installed at the positions of FC4 and FC6, along with multiple IRIS diaphragms.~\cite{early-CRIS-PKU}. It is noteworthy that during routine optimization procedures, adjustments to the laser beam alignment are typically minimal, as the optimal condition is largely maintained once the lasers are introduced into the beamline. This stability can be partially attributed to the anti-collinear geometry, which significantly reduces the distance between the laser entrance window and IR. 
\par Under these conditions, HFS spectra of $^{85,87}$Rb were measured using both laser-frequency scan and voltage scan. During the laser-frequency scan, the voltage scanner is grounded, and the frequency of the first-step laser is tuned to match the Doppler-shifted D2 transition. In the voltage scan, the first-step laser is locked to a fixed frequency via WM1. By tuning the voltage (up to $\pm$ 500 V) applied to the voltage scanner, the beam velocity is scanned across the Doppler-shift HFS structure. The total beam energy is determined by recording the voltages applied to the RFQ and the voltage scanner, using a Keysight 34470A multimeter in conjunction with KV-30A and KV-10A voltage dividers, respectively. Laser resonantly ionized ions are detected by the MagneTOF as the function of laser frequency or tuning voltage~\cite{CLS-DAQ}, enabling the measurement of the HFS spectra. 

\par Figure~\ref{Fig2}~(a) shows the HFS spectra of $^{85,87}$Rb measured using the laser-frequency scan, which are fitted using a $\chi^2$-minimization routine in SATLAS~\cite{Satlas2018}. The notable asymmetry in the resonance peaks of the HFS spectra is found to be strongly associated with the helium buffer gas pressure injected into the RFQ. The spectra with blue lines correspond to measurements with a spectral resolution of approximately 100~MHz and an overall experimental efficiency of around 1:200 estimated by detecting the RIS counts and ion counts at FC2. However, due to the extremely low beam intensity, direct beam current measurement using FC2 is challenging. Thus, the beam intensity at FC2 is calibrated prior to the HFS measurement by chopping the ion beam using a beam gate~\cite{RFQ-PKU} and detecting the corresponding ion counts using MagneTOF, taking into account the total transmission efficiency and the natural abundance (72.17\%) of $^{85}$Rb. As a result, the highest RIS count rate of about 1500 in the $^{85}$Rb HFS spectrum is obtained with a total beam current of 0.05 pA at FC2. This translates to a total experimental efficiency of 1:200, accounting for ion transmission, neutralization, resonance ionization and ion detection efficiencies, but excluding the RFQ transmission efficiency.
\par We also performed HFS spectra measurements using higher laser power, achieving an efficiency better than 1:50; however, this came at the cost of a spectral resolution limited to approximately 1.5 GHz due to laser power broadening. Further improvement in spectral resolution down to 30~MHz can be easily achieved by reducing the laser power and introducing timing delays for the second- and third-step laser pulses, although this results in lower total efficiency~\cite{CRISNIMB-2020-2}. For example, as shown by red lines in Fig.~\ref{Fig2}~(a), a spectral resolution of 50~MHz was achieved by reducing the laser power alone, albeit with a reduced overall efficiency of 1:500.
These measurements are independently repeated using the voltage scan, leading to the same conclusions regarding efficiency and spectral resolution. With the 30-MHz spectral resolution, we can quantitatively estimate the energy spread of the ion beam from the RFQ. After subtracting the contribution from natural linewidth (approximately 6~MHz) and the linewidth of the injection-seeded laser (at least 10~MHz~\cite{inseeded2018}), the remaining width attributed to Doppler broadening corresponds to a maximum energy spread of $\sim$3.6~eV, assuming no additional laser power broadening.
\par Although the HFS spectra in Fig.~\ref{Fig2}~(a) were measured with different spectral resolution and exhibit slightly asymmetry, the extracted magnetic and quadrupole HFS parameters ($A_{\rm u}$, $A_{\rm l}$ and $B_{\rm u}$) for upper $5p$ $^{2\!}P^{\rm{o}}_{3/2}$ and lower $5s$ $^{2\!}S_{1/2}$ states, as well as the IS for $^{85,87}$Rb are consistent with each other and agree well with the literature values~\cite{Rb1981}, as illustrated in Fig.~\ref{Fig2}~(b). We further compare the atomic parameters obtained from both laser-frequency scan and voltage scan, demonstrating excellent agreement between these two approaches. Overall, the performance achieved with the PLASEN setup is comparable to the current state-of-the-art CRIS setup~\cite{Fr-2013,Cu2017}. 
\par With this successful commissioning experiment, we will soon install the system at BRIF to perform laser spectroscopy measurements of neutron-rich Rb isotopes. Therefore, it is critical to assess the potentially achievable measurement sensitivity, as the targeted most neutron-rich $^{100}$Rb is produced with a yield as low as 100 pps at BRIF. Based on an efficiency of 1:200, a RFQ transmission efficiency of 50\%, and a signal-to-noise ratio exceeding 150 (Fig.~\ref{Fig2}(a)), simulations indicate that with an ion beam intensity of 100 pps, a full HFS spectrum of the Rb isotope can be obtained with a five-sigma significance on the smallest resonance peak within two hours. It is worth noting that the isobaric contaminants ($A=100$ Sr and Y) are expected to be produced at rates 100 times higher than $^{100}$Rb inside the target irradiated by the 100-MeV proton beam at BRIF. Therefore, in the worst-case scenario—assuming the same release and ionization efficiencies from the ISOL target and equal ion background contributions from the Sr and Y isobars as from the Rb isotopes—the HFS spectrum of the Rb isotope can still be measured with five-sigma significance within ten hours.

In summary, through the implementation of the RFQ cooler-buncher and the multistep laser resonance ionization technique, we have successfully established a high-resolution and high-sensitivity collinear resonance ionization laser spectroscopy system-PLASEN. The setup was commissioned by probing the HFS spectra of $^{85,87}$Rb isotopes, achieving an overall efficiency of 1:200 with a spectral resolution of 100~MHz. This result yields an experimental sensitivity sufficient for laser spectroscopy measurements with beam intensity of 100 pps, matching the general performance level of the CRIS setup at CERN-ISOLDE. The extracted HFS constants and IS are in line with the literature values, demonstrating the reliability of the PLASEN setup. 
\par The performance of the PLASEN system opens the possibilities for laser spectroscopy measurements of unstable isotopes at BRIF and other upcoming RIB facilities in China. Studies of the nuclear properties of neutron-rich Rb and Cs isotopes are expected to be conducted at BRIF in the near future. In particular, Rb isotopes beyond $N=60$ and Cs isotopes beyond $N=90$ are of special interest, as they are anticipated to exhibit pronounced nuclear deformation. With ongoing improvement of the BRIF and the construction of the HIAF facility~\cite{HIAF2022}, high-resolution laser spectroscopy measurements of even more exotic isotopes can be anticipated in the near future.
\vspace{-4mm}
\section*{Acknowledgments}
\vspace{-2mm}
This work was supported by National Natural Science Foundation of China (Nos.12027809, 12350007, 12305122), National Key R\&D Program of China (Nos.~2023YFE0101600, 2023YFA1606403, 2022YFA1605100). The authors would like to thank the members of the CRIS and COLLAPS collaboration for their valuable discussion and support in the development of the setup. 
\vspace{-2mm}
\bibliographystyle{elsarticle-num}
\vspace{-2mm}
\bibliography{CRIS-PKU}

\providecommand{\noopsort}[1]{}\providecommand{\singleletter}[1]{#1}%
\begin{thebibliography}{10}
\expandafter\ifx\csname url\endcsname\relax
  \def\url#1{\texttt{#1}}\fi
\expandafter\ifx\csname urlprefix\endcsname\relax\def\urlprefix{URL }\fi
\expandafter\ifx\csname href\endcsname\relax
  \def\href#1#2{#2} \def\path#1{#1}\fi

\bibitem{NRP2025}
Y.~L. Ye, et~al., Nat.~Rev.~Phys. 7 (2025) 21--37.

\bibitem{PPNP2023}
X.~F. Yang, et~al., Prog.~Part.~Nucl.~Phys. 129 (2023) 104005.

\bibitem{COLLAPS2017}
R.~Neugart, et~al., J. Phys. G: Nucl. Part. Phys. 44 (2017) 064002.

\bibitem{roc}
L.~Vermeeren, et~al., Phys. Rev. Lett. 68 (1992) 1679.

\bibitem{CRIS-NIMB2020}
A.~Vernon, et~al., Nucl. Instrum. Methods Phys.~Res.~B 463 (2020) 384.

\bibitem{MIRACLS}
V.~Lagaki, et~al., Nucl. Instrum. Methods Phys.~Res.~A 1014 (2021) 165663.

\bibitem{Ac-moment2017}
R.~Ferrer, et~al., Nat.~Commun. 8 (2017) 14520.

\bibitem{RADRIS2023}
S.~Raeder, et~al., Nucl. Instrum. Methods Phys. Res.~B 541 (2023) 370.

\bibitem{Cu2017}
R.~P. de~Groote, et~al., Phys. Rev. C 96 (2017) 041302.

\bibitem{Fr-PRX}
K.~M. Lynch, et~al., Phys. Rev. X 4 (2014) 011055.

\bibitem{RaF}
R.~F. {Garcia Ruiz}, et~al., Nature 581 (2020) 396--400.

\bibitem{BRIF2020}
T.~J. Zhang, et~al., Nucl. Instrum. Methods Phys.~Res.~B 463 (2020) 123.

\bibitem{HIAF2022}
X.~H. Zhou, et~al., AAPPS Bull. 32 (2022) 35.

\bibitem{CLS-PKU}
S.~W. Bai, et~al., Nucl. Sci. Tech. 33 (2022) 9.

\bibitem{CLS-BRIF}
S.~J. Wang, et~al., Nucl. Instrum. Methods Phys. Res. Sect. A 1032 (2022) 166622.

\bibitem{early-CRIS-PKU}
P.~Zhang, et~al., Nucl.~Instrum.~Methods Phys.~Res. Sect.~B 541 (2023) 37.

\bibitem{RFQ-PKU}
Y.~S. Liu, et~al. (2025).
\newblock \href {http://arxiv.org/abs/2502.10740} {\path{arXiv:2502.10740}}.

\bibitem{Rb1981}
C.~Thibault, et~al., Phys. Rev. C 23 (1981) 2720--2729.

\bibitem{inseeded2018}
M.~Reponen, et~al., Nucl. Instrum. Methods Phys. Res. Sect. A 908 (2018) 236--243.

\bibitem{CLS-DAQ}
Y.~C. Liu, et~al., Nucl. Sci. Tech. 34 (2023) 38.

\bibitem{Satlas2018}
W.~Gins, et~al., Comput. Phys. Comm. 222 (2018) 286--294.

\bibitem{CRISNIMB-2020-2}
A.~Koszorus, et~al., Nucl. Instrum. Methods Phys. Res.~B 463 (2020) 398.

\bibitem{Fr-2013}
K.~T. Flanagan, et~al., Phys. Rev. Lett. 111 (2013) 212501.

\end{thebibliography}

\end{document}